\documentclass[12pt]{iopart}

\selectfont 

\usepackage{graphicx}       
\usepackage{dcolumn}        
\usepackage{bm}             
\usepackage{amssymb}
\usepackage{amstext}

\newcommand{\nn}{\nonumber}

\newcommand{\sig}{\sigma}

\newcommand{\eps}{\epsilon}

\newcommand{\rstar}{r_\ast}
\newcommand{\astar}{{a_\ast}}
\newcommand{\aw}{z}

\newcommand{\Rinc}{B^{\text{(inc)}}_{lm\omega}}
\newcommand{\Rref}{B^{\text{(refl)}}_{lm\omega}}
\newcommand{\Rtrans}{B^{\text{(trans)}}_{lm\omega}} 

\newcommand{\Yspher}{{}_{s}Y}

\newcommand{\diffop}{ \hat{\mathcal{L}}_x }
\newcommand{\diffopm}{ \hat{\mathcal{L}}_{-x} }
\newcommand{\alpco}{\alpha_{lm}^{(2)}}
\newcommand{\betco}{\beta_{lm}^{(2)}}

\begin{document}

 \title{Polarization of Long-Wavelength Gravitational Waves by Rotating Black Holes}

\author{Sam R. Dolan}
 \address{%
 School of Mathematical Sciences, University College Dublin, Belfield, Dublin 4, Ireland
}%
 \ead{sam.dolan@ucd.ie}

\date{\today}

\begin{abstract}
The scattering cross section for a long-wavelength planar gravitational wave impinging upon a rotating black hole is calculated, for the special case in which the direction of incidence is aligned with the rotation axis. We show that black hole rotation leads to a term in the cross section that is proportional to $a\omega$. Hence, contrary to some claims, co-rotating and counter-rotating helicities are scattered differently, and  a partial polarization is induced in an unpolarized incident wave.

The scattering amplitudes are found via partial wave series. To compute the series, two ingredients are required: phase shifts and spin-weighted spheroidal harmonics. We show that the phase shifts may be found from low-frequency solutions of the radial Teukolsky equation derived by Mano, Suzuki and Takasugi. The spheroidal harmonics may be expanded in spherical harmonics; we present expansions accurate to second order in $a \omega$. The two ingredients are combined to give explicit expressions for the helicity-conserving and helicity-reversing amplitudes, valid in the long-wavelength limit. 
\end{abstract}

\maketitle

\section{\label{sec:introduction}Introduction}

Gravitational waves (GWs) are propagating ripples in space-time whose existence is predicted by General Relativity. There is strong indirect evidence for their existence, for example, from thirty-five years of pulsar timing measurements \cite{Taylor-1994}. Yet due to the tiny expected amplitude of waves reaching Earth, GWs have not been measured directly. Now, nine decades after the formulation of Einstein's theory, many experimentalists are optimistic that ``first light'' detections are imminent, at either (existing) ground-based \cite{Waldman-2006} or (future) space-based \cite{Danzmann-2003} interferometers.

GWs are of interest to astronomers because they are generated by some of the most energetic astrophysical processes, such as binary mergers, supernovae and galaxy collisions. Electromagnetic radiation carries relatively little information about the most energetic regions of such processes, because photons are strongly scattered, absorbed and thermalized by intervening matter. On the other hand, gravitational waves are only weakly coupled to matter, and carry information about the dynamics at the heart of such processes. 



In this note, we are motivated by a simple general question: does the rotation of matter induce a polarization in a gravitational wave? To explore this question, we study a special case. We compute the differential scattering cross section for a long-wavelength planar gravitational wave impinging along the rotation axis of a black hole. We will assume that the incident wave is monochromatic, long-lasting, and sufficiently weak that the gravitational field equations may be linearized. The scenario is then characterized by just two dimensionless parameters,
\begin{equation}
M |\omega| = \pi r_s / \lambda ,  \quad \quad \text{ and } \quad \quad \astar = a / M = J / M^2
\end{equation}
(with units $G = c = 1$, used throughout). The coupling $M |\omega|$ expresses the ratio of Schwarzschild horizon size $r_s $ to incident wavelength $\lambda$, and $0 \le \astar < 1$ is a measure of the rotation rate of hole. 
In this paper, we concern ourselves only with the long-wavelength regime, in which $M |\omega| \ll 1$. 

The frequency $\omega$ may take either sign, depending on the helicity of the incident wave. Positive $\omega > 0$ corresponds to a circularly-polarized incident wave co-rotating with the black hole, whereas negative $\omega < 0$ corresponds to a counter-rotating helicity. If the scattering interaction is able to distinguish between these cases, then a polarization will result.


Over the years, various authors \cite{Westervelt-1971,Peters-1976,Sanchez-1976,Matzner-1977,DeLogi-1977,Futterman-1988,Doran-2002,Dolan-2008} have shown that, for a non-rotating massive body ($\astar = 0$) in the long wavelength limit ($M\omega \ll 1$) the cross section depends on the spin $s$ of the scattered field as follows: 
\begin{eqnarray}
\lim_{M |\omega| \rightarrow 0} \, \left( \frac{1}{M^2} \, \frac{d \sigma}{d \Omega} \right)  &\approx  
\left\{ 
\begin{array}{llll} 
  \frac{1}{\sin^4(\theta / 2)}  & \quad s = 0, & \quad \text{Scalar wave} & \text{[a]} \\ 
  \frac{\cos^2(\theta /2)}{\sin^4(\theta / 2)}  & \quad s = \frac{1}{2}, & \quad \text{Neutrino} & \text{[b]}  \\
  \frac{\cos^4(\theta /2)}{\sin^4(\theta / 2)}  & \quad s = 1, & \quad \text{Photon}  & \text{[c]}   \\
  \frac{\cos^8(\theta /2) \, + \, \sin^8(\theta / 2)}{\sin^4(\theta / 2)} & \quad s = 2, & \quad \text{Grav. wave} & \text{[d].} 
\end{array} \right.
\label{csec-low-approx}
\end{eqnarray}
The gravitational result is somewhat anomalous, in that it doesn't follow the same general rule $\left[ d\sig/d\Omega = M^2 \cos^{4s}(\theta/2) / \sin^4(\theta/2) \right]$ as the other fields. This is related to the fact that the helicity of a gravitational wave is not conserved by the scattering process \cite{Matzner-1977, DeLogi-1977, Dolan-2008}.

The influence of rotation on the scattering of low-frequency waves ($M|\omega| \ll 1$) has been considered by a number of authors \cite{Mashhoon-1973, Mashhoon-1974, Mashhoon-1975, Guadagnini-2002, Barbieri-2004, Barbieri-2005, DeLogi-1977, Matzner-1977, Guadagnini-2008}. It seems that no clear consensus has yet emerged. Considering electromagnetic radiation, Mashhoon \cite{Mashhoon-1973} noted that ``one must expect partial polarization in the scattered light when an unpolarized wave is incident on a Kerr black hole", and went on to observe that ``the polarizing property of a Kerr black hole is probably maintained for very low frequencies". However, the corollary that incoming gravitational waves will also be polarized at low frequencies was thrown into doubt by a study \cite{DeLogi-1977} in which the scattering amplitude was computed using Feynman-diagram techniques. The authors concluded that ``the angular momentum of the scatterer has no polarizing effect on incident, unpolarized gravitational waves" (even though they found that unpolarized electromagnetic waves \emph{were} polarized). Some doubts about the gauge-invariance of the results in \cite{DeLogi-1977} have been raised. To clarify the issue, the Feynman-diagram approach was recently revisited and improved in \cite{Guadagnini-2008}. In a series of papers, Guadagnini and Barbieri \cite{Guadagnini-2002, Barbieri-2004, Barbieri-2005, Guadagnini-2008} have argued that the net polarisation $\mathcal{P}$ induced by rotating classical matter is
\begin{equation}
\mathcal{P} \equiv \frac{ {\frac{ d\sig}{ d\Omega }}{(\omega>0)} - {\frac{d\sig}{d\Omega}}{(\omega<0)} }
{{\frac{ d\sig}{ d\Omega }}{(\omega>0)} + {\frac{d\sig}{d\Omega}}{(\omega<0)}   }  \approx  - |s| \omega (J / M) \theta^2   \label{pol-guadagnini}
\end{equation}
where $\theta$ is the scattering angle, which is assumed to be small.

In this paper we combine the partial wave approach of Matzner \emph{et al.} \cite{Chrzanowski-1976, Matzner-1977, Matzner-1978, Futterman-1988} with the low-frequency asymptotics of Mano, Suzuki and Takasugi \cite{MST} to derive the lowest-order correction to cross section (\ref{csec-low-approx})[d] for a gravitational wave impinging along the rotation axis of a Kerr hole. We find a term proportional to $a \omega$ that couples the helicity of the incident wave with the spin of the hole. The total cross section is given in Eq. (\ref{csec-final}). The resulting polarization, given in Eq. (\ref{pol-final}), is in qualitative agreement with (\ref{pol-guadagnini}) at small angles, but differs by a factor of two.  

The remainder of this paper is organised as follows. In section \ref{sec-partial} we briefly recap the results of Matzner \emph{et al.} \cite{Chrzanowski-1976, Matzner-1977, Matzner-1978, Futterman-1988} to write the cross section in terms of amplitudes which are expressed as partial wave series. In \ref{sec-phaseshifts}, we define the phase shifts and discuss their asymptotic values \cite{MST,MT} in the long-wavelength regime ($M |\omega| \ll 1$). In \ref{sec-spheroidal} we expand the relevant spin-weighted spheroidal harmonics to second order in $a\omega$. In \ref{subsec-amplitudes}, we identify that part of the scattering amplitudes that is proportional to $a \omega$, and compute the corresponding contributions to the cross section. We conclude with a brief discussion in section \ref{sec:discussion}. 

\section{Analysis}
\subsection{\label{sec-partial}Partial Wave Series}
In the late 1970s, Matzner and co-workers \cite{Chrzanowski-1976, Matzner-1977, Matzner-1978, Handler-1980} showed that the differential cross section for the scattering of gravitational waves by a rotating black hole can be written as the sum of the square magnitude of two amplitudes,
\begin{equation}
\frac{d \sigma}{d \Omega} = |f(\theta)|^2 + |g(\theta)|^2. \label{csec-def}
\end{equation}
where $\theta$ is the scattering angle. 
In the special case in which the incident wave vector is parallel to the rotation axis, the amplitudes may be expressed as the following partial-wave series,
\begin{eqnarray}
f( \theta ) &= \frac{\pi}{i \omega} \sum_{P=\pm 1} \sum_{l=2}^\infty \left[ \exp(2i \delta_{l2\omega}^P) - 1 \right] \,  {}_{-2}S_l^2(0; a \omega) \,\, {}_{-2}S_l^2(\theta; a \omega) ,  \label{fg1}  \\
g( \theta ) &= \frac{\pi}{i \omega} \sum_{P=\pm 1} \sum_{l=2}^\infty P (-1)^l \left[ \exp(2i \delta_{l2\omega}^P) - 1 \right] \,  {}_{-2}S_l^2(0; a \omega) \,\, {}_{-2}S_l^2(\pi - \theta; a \omega) . \label{fg2}
\end{eqnarray}
In these expressions, $\exp(2i \delta_{lm\omega}^P)$ are phase factors to be determined from a radial equation, ${}_sS_l^m(\theta; a \omega)$ are spin-weighted spheroidal harmonics. 
Note the presence of the sum over even and odd parities, $P = \pm 1$. 

The first amplitude $f(x)$ corresponds to (that part of) the interaction which preserves the helicity (i.e. for which the helicity of the scattered wave is the same as the helicity of the incident wave). The second amplitude $g(x)$ corresponds to (that part of) the interaction which reverses the incident helicity. The helicity-reversing amplitude is non-zero because the phase shifts $\delta_{lm\omega}^P$ depend on parity $P$. In this respect, gravitational wave scattering is unlike scalar, neutrino, or electromagnetic scattering.

In \cite{Dolan-2008} it was shown that, to lowest order in $M\omega$, the Schwarzschild ($a=0$) amplitudes may be written as
\begin{equation}
f_{\text{Schw}} (x) = M e^{i \Phi} \frac{\Gamma(1-i\epsilon)}{\Gamma(1 + i\epsilon)} \frac{ [\frac{1}{2} (1+x)]^2 }{[ \frac{1}{2} (1 - x)]^{1-i\epsilon}}
\end{equation}
and
\begin{equation}
g_{\text{Schw}} (x) = M e^{i \Phi} \frac{\Gamma(1-i\epsilon)}{\Gamma(1 + i\epsilon)} [\frac{1}{2} (1-x)]
\end{equation}
where $\Phi = 2\eps \ln | 2\eps |$, $x = \cos\theta$, $\eps = 2M\omega$ and $\Gamma(z)$ is the Gamma function. In the following sections, we show that black hole rotation ($a > 0$) introduces a term in each amplitude which is proportional to $a \omega$. 

\subsection{\label{sec-phaseshifts}Phase Shifts}
The phase shifts are found from solutions ${}_sR_{lm\omega}(r)$ to Teukolsky's \cite{Teukolsky-1972} radial equation that satisfy the following boundary conditions,
\begin{equation}
{}_{s}R_{lm\omega}(r) \sim \left\{ \begin{array}{l l} \Rtrans \Delta^{-s} e^{- i k \rstar} & \rstar \rightarrow -\infty  \\ \Rinc r^{-1} e^{-i\omega \rstar}  + \Rref r^{-(2s+1)} e^{+i\omega \rstar}  & \rstar \rightarrow +\infty  \end{array} \right.  .
\label{teuk-bc}
\end{equation}
Here, $\Delta = r^2 - 2Mr + a^2$, $k = \omega - \astar m / [2(1+\sqrt{1-\astar^2})]$, $\rstar$ is the standard tortoise coordinate, $s = -2$ for a gravitational wave and $\Rinc$, $\Rref$ and $\Rtrans$ are complex constants. The phase shifts are found from
\begin{equation}
\exp\left( 2 i \delta_{lm\omega}^P \right) = (-1)^{l+1}  \left( \frac{\text{Re}(C) + 12iM\omega P}{16 \omega^4} \right) \frac{\Rref}{\Rinc}  \label{eq-phaseshift1}
\end{equation}
where $C$ is the Starobinskii constant,
\begin{eqnarray}
\left[\text{Re}(C)\right]^2 = & ((\lambda+2)^2 + 4am\omega - 4a^2 \omega^2) \left[\lambda^2 + 36am\omega - 36a^2 \omega^2 \right] \nn \\ & \quad + (2\lambda + 3)(96a^2\omega^2 - 48a\omega m) - 144 \omega^2 a^2, 
\end{eqnarray}
and $\lambda = A_{lm} - 2 m a \omega + a^2 \omega^2$ is the angular separation constant. 

Low-frequency analytic results for the phase shifts may be obtained via the formalism developed by  
Mano, Suzuki and Takasugi \cite{MST, MT} (MST). In the MST approach, reviewed in \cite{Sasaki-Tagoshi}, solutions to the Teukolsky radial equation satisfying boundary conditions (\ref{teuk-bc}) are expressed as infinite series of special functions. Two series are used. The `horizon' series of ${}_2F_1$ hypergeometric functions is convergent at all radii up to (but not including) spatial infinity. The `far-field' series of Coulomb wavefunctions is convergent up to (but not including) the outer horizon. The complex constants  $\Rinc$, $\Rref$ and $\Rtrans$ are determined by matching the two series. It is found that
\begin{eqnarray}
\Rinc &= A_s e^{i \epsilon \kappa} \omega^{-1} \left[ K_\nu(s) - i e^{i \pi \nu} \frac{\sin \left(\pi(\nu-s+i\eps) \right)}{\sin \left(\pi(\nu+s-i\eps) \right)} K_{-\nu-1}(s) \right] A_+^\nu , \\
\Rref &= A_s e^{i \epsilon \kappa} \omega^{-1-2s} \left[  K_\nu(s)  + i e^{i \pi \nu} K_{-\nu-1}(s) \right] A_-^\nu .
\end{eqnarray}
Here $A_s$ is just a normalization constant, and we follow the conventions of \cite{MST} by defining
\begin{equation}
\eps = 2M\omega \quad \quad \text{ and  } \quad \quad \kappa = \sqrt{1- \astar^2}.
\end{equation}
Note that $\epsilon$ must be positive in the MST expressions; the $\omega < 0$ results may be found via the symmetry of the radial function ${}_sR_{lm\omega} = {}_sR_{l-m-\omega}^\ast$ from which it follows that $B^{\text{(inc/refl)}}_{lm -\omega} = B^{\text{(inc/refl)} \ast}_{l -m \omega}$.

The parameter $\nu$ is known as the ``renormalized angular momentum'' and has the low-frequency expansion
\begin{equation}
\fl  \nu = l  +  \frac{\eps^2 }{2l + 1} \left[ -2 - \frac{s^2}{l (l+1)} + \frac{\left[ (l+1)^2 - s^2 \right]^2}{(2l+1)(2l+3)(2l+3)} - \frac{(l^2 - s^2)^2}{(2l-1)(2l)(2l+1)} \right] + \mathcal{O}(\eps^3).
\end{equation}
Note the absence of a linear term in $\epsilon$. 

The coefficients $K_\nu$ and $K_{-\nu-1}$ may be computed via a complicated series expansion, detailed in \cite{MST}. This is not necessary for our purposes, because $  K_{-\nu - 1} / K_{\nu}  \sim \mathcal{O}(\epsilon^{2l-1}) $. Hence $  K_{-\nu - 1}$ may be neglected, and the $K_\nu$ terms will cancel upon taking the ratio of $\Rinc$ and $\Rref$.

The coefficients $A_+^\nu$ and $A_-^\nu$ are given by 
\begin{eqnarray}
A_+^{\nu} &= 2^{-1+s} \eps^{-i\eps} e^{i (\pi/2) (\nu + 1 - s)} e^{-\pi \eps / 2} \frac{\Gamma(\nu + 1 - s + i\eps)}{\Gamma(\nu + 1 + s -i\eps)} \sum_{n=-\infty}^{\infty} a_n^\nu (s) ,  \label{Ap-eq} \\
A_-^{\nu} &= 2^{-1-s} \eps^{+i \eps} e^{-i (\pi/2) (\nu + 1 + s)} e^{-\pi \eps / 2} \sum_{n=-\infty}^{\infty} (-1)^n \frac{ (\nu + 1 + s - i\eps)_n }{ (\nu + 1 -s + i \eps)_n } a_n^\nu (s), 
\label{Am-eq}
\end{eqnarray}
where $a_n^\nu \sim \mathcal{O}( \eps^{|n|} )$. Here, we wish to conduct an expansion accurate to second order in $M\omega$; hence we require explicit formulae for $a_{-2}^\nu,  \ldots, a_{2}^\nu$. These were computed in \cite{MST} and are listed in \ref{appendix-mst}. 

After inserting equations (\ref{a-nu-m2}--\ref{a-nu-2}) into (\ref{eq-phaseshift1}), and taking some care with series expansions in the small parameter $\eps$, we find
\begin{eqnarray}
\fl\quad\quad  \exp({2 i \delta_{lm\omega}^-}) = e^{2i\eps \ln 2\eps} e^{-i \eps \kappa} e^{- i \pi (\nu - l)} \frac{\Gamma(l + 1 - i\eps)}{\Gamma(l + 1 + i\eps)} e^{4 i \eps / l(l+1)} \left[ 1 +  \alpco \eps^2  + \mathcal{O}(\eps^3) \right]  \label{phase-shift-negative}
\end{eqnarray}
for positive $\omega$. 
Note that there is no first-order term in the square brackets; the linear terms cancel exactly. The second-order coefficient is
\begin{equation}
\alpco = - \frac{i m \astar}{l(l+1)} - \frac{ 12 i m \astar }{(l-1) l^2 (l+1)^2 (l+2)} ,
\end{equation}
where $m$ is the azimuthal number. Note that the $e^{-i \pi (\nu - l)}$ and $e^{4 i \eps / l(l+1)}$ factors also give an $l$-dependent contribution at second order in $\epsilon^2$. 

Via the symmetry $B^{\text{(ref/inc)}}_{lm -\omega} = B^{\text{(ref/inc)} \ast}_{l -m \omega}$ it follows that $\exp(2i \delta^{-}_{l m -\omega} ) = \exp(2i \delta^{-}_{l -m \omega})^\ast$. Hence the phase shift may be written more generally as 
\begin{eqnarray}
\fl\quad\quad \exp({2 i \delta_{lm\omega} ^-}) = e^{2i\eps \ln |2\eps|} e^{-i \eps \kappa} e^{- i \, \pi (\nu - l) \omega / |\omega| } \frac{\Gamma(l + 1 - i\eps)}{\Gamma(l + 1 + i\eps)} e^{4 i \eps / l(l+1)} \left[ 1 +  \alpco \eps^2  + \mathcal{O}(\eps^3) \right]  .  \label{phase-shift-negative-both}
\end{eqnarray}
This expression is valid for either sign of $\eps = 2M\omega$. 

In a previous study \cite{Dolan-2008}, the low-$M\omega$ approximations of Poisson and Sasaki \cite{Poisson-1995} were used to show that the Schwarzschild ($\astar = 0$) phase shifts are
\begin{eqnarray}
\exp( 2 i \delta_{lm\omega}^- )= e^{2i\eps \ln 2\eps} e^{-i \eps} \frac{\Gamma(l + 1 - i\eps)}{\Gamma(l + 1 + i\eps)} e^{4 i \eps / l(l+1)} \left[ 1 +  \mathcal{O}(\eps^2) \right] .  \label{schw-phase}
\end{eqnarray}
This is consistent with (\ref{phase-shift-negative}) since $\kappa = 1$ in the non-rotating case. Moreover, it is remarkable that, when $\astar = 0$, equation (\ref{schw-phase}) actually holds to one order higher in $\eps$ than first supposed. 

\subsection{Spheroidal Harmonics\label{sec-spheroidal}}
The spin-weighted spheroidal harmonics ${}_sS_l^m(\theta; \aw = a\omega)$ are solutions of the equation
\begin{eqnarray}
\fl\frac{1}{\sin \theta} \frac{d}{d\theta} \left(\sin \theta \frac{d S}{d \theta} \right) + & \left( \aw^2 \cos^2 \theta - \frac{m^2}{\sin^2 \theta} - \frac{2 m s \cos \theta}{\sin^2 \theta} 
-2 \aw s \cos \theta - s^2 \cot^2 \theta + s + A_{lm} \right) S = 0 \label{angular-eq}
\end{eqnarray}
where the spheroidicity parameter is
\begin{equation}
\aw = a \omega .
\end{equation}
These functions are regular at $\theta = 0$ and $\theta = \pi$, and are normalized so that
\begin{eqnarray}
\int_{-1}^{1} d(\cos \theta) {}_s S_l^m(\theta; \aw) {}_s S_j^m(\theta; \aw) = \frac{ \delta_{lj} }{2 \pi} .
\end{eqnarray}
Note that we suppress the azimuthal factor $e^{im\phi}$ throughout.

The spheroidal harmonics may be decomposed into a sum over spherical harmonics $\Yspher_{j}^m(\theta)$ of the same spin $s$ and azimuthal number $m$. That is,
\begin{equation}
{}_{s} S_l^m(\theta; \aw) = \sum_{j = \text{max}(|m|,|s|)}^\infty b^{(l)}_{j} \, \Yspher_{j}^m(\theta) ,  \label{S-expansion}  
\end{equation}
The expansion coefficients $b_j^{(l)}$ satisfy a five-term recurrence relation, 
\begin{eqnarray}
\fl \quad \left( \aw^2 c^{(2)}_{k, k-2} \right)  b_{k-2}^{(l)} + \left(\aw^2 \omega^2 c^{(2)}_{k, k-1} - 2 \aw s c^{(1)}_{k, k-1} \right) b_{k-1}^{(l)}  + \left( \aw^2 c_{kk}^{(2)} - 2 \aw s c_{kk}^{(1)} - k(k+1) \right) b_{k}^{(l)}  \nn \\
\fl \quad\quad\quad
+ \, \left( \aw^2 c^{(2)}_{k, k+1} - 2 \aw s c^{(1)}_{k, k+1} \right) b_{k+1}^{(l)}  + \left( \aw^2 c^{(2)}_{k, k+2} \right) b_{k+2}^{(l)}  = - E_{lm} b_{k}^{(l)}  \label{eq-quindiag}
\end{eqnarray}
where $E_{lm} = A_{lm} + s(s+1)$ and
\begin{eqnarray}
\fl c_{k j}^{(1)} = \int d\Omega \, {}_sY_k^m (\theta) \cos \theta \, {}_sY_j^m(\theta)  &= \sqrt{\frac{2j+1}{2k+1}} \left<j, 1, m, 0| k, m\right>\left< j, 1, -s, 0 | k, -s \right> , \\
\fl c_{k j}^{(2)} = \int d\Omega \, {}_{s}Y_k^m (\theta) \cos^2 \theta  \, {}_{s}Y_j^m(\theta) &= \frac{1}{3} \delta_{kj} + \frac{2}{3} \sqrt{\frac{2j+1}{2k+1}} \left< j,2,m,0 | k,m \right> \left< j, 2, -s, 0 | k, -s \right>  .
\end{eqnarray}
The numbers $\left< j_1, j_2, m_1 , m_2 | j, m \right>$ are Clebsch-Gordan coefficients. For more details see, for example, Appendix A in \cite{Hughes-2000}.

We wish to expand the spheroidal harmonics to second order in $z$. Noting that $b_{l \pm n}^{(l)} \sim \mathcal{O}( \aw^n )$, let us make the expansion
\begin{eqnarray}
b_{l-2}^{(l)} &=  \aw^2 d_{-2}^{(0)} ,  \\
b_{l-1}^{(l)} &=  \aw d_{-1}^{(0)} + \aw^2 d_{-1}^{(1)} ,  \\
b_{l}^{(l)}    &= 1 + \aw^2 d_{0}^{(2)} ,   \\ 
b_{l+1}^{(l)} &=  \aw d_{+1}^{(0)} + \aw^2 d_{+1}^{(1)} , \\ 
b_{l+2}^{(l)} &=  \aw^2 d_{+2}^{(0)} .  \label{bdef}
\end{eqnarray}
The normalisation condition implies that $(d_0)^{2} = -\frac{1}{2}\left[ (d_{-1}^{(0)} )^2 + (d_{+1}^{(0)})^2 \right]$. The remaining six unknowns $\{ d_{-2}^{(0)},  d_{-1}^{(0)}, d_{-1}^{(1)}, d_{0}^{(2)}, d_{+1}^{(0)}, d_{+1}^{(1)}, d_{+2}^{(0)}  \}$ are determined from the equations 
\begin{eqnarray}
(l+1) d_{+1}^{(0)} &= - s c_{l+1, l}^{(1)} ,  \label{deqs1} \\
l d_{-1}^{(0)} &=  s c_{l-1, l}^{(1)},   \\
2(l+1) d_{+1}^{(1)} &= (-2s c_{l+1,l+1}^{(1)} + E^{(1)}_{lm}) d_{+1}^{(0)} + c_{l+1,l}^{(2)} , \\
-2l d_{-1}^{(1)} &= (-2s c_{l-1,l-1}^{(1)} + E^{(1)}_{lm} ) d_{-1}^{(0)} + c_{l-1,l}^{(2)} , \\ 
2(2l+3) d_{+2}^{(0)} &= c_{l+2,l}^{(2)} - 2s c_{l+2, l+1}^{(1)} d_{+1}^{(0)} , \\
-2(2l-1) d_{-2}^{(0)} &= c_{l-2,l}^{(2)}  - 2s c_{l-2, l-1}^{(1)} d_{-1}^{(0)} ,   \label{deqs2}
\end{eqnarray}
where the angular eigenvalue has been expanded in powers of $\aw = a \omega$ as
\begin{equation}
E_{lm}  =  l (l + 1) + \sum_{k = 1}^\infty E_{lm}^{(k)} \aw^k 
\end{equation}
and the first few coefficients are determined by the identities
\begin{eqnarray}
E_{lm}^{(1)} - 2 s c_{ll}^{(1)} &= 0 \, , \\
E_{lm}^{(2)} + c_{ll}^{(2)} - 2s c_{l,l-1}^{(1)} d_{-1}^{(0)} - 2s c_{l, l+1}^{(1)} d_{+1}^{(0)} &= 0 \, , \\
E_{lm}^{(3)} + c_{l,l-1}^{(2)} d_{-1}^{(0)} + c_{l,l+1}^{(2)} d_{+1}^{0} - 2s c_{l,l-1}^{(1)} d_{-1}^{(1)} - 2s c_{l,l+1}^{(1)} d_{+1}^{(1)} &= 0 .
\end{eqnarray}

For this calculation, we need only the $m = 2$, $s = -2$ harmonics. The required Clebsch-Gordan coefficients are listed in \ref{appendix-mst}, (\ref{cleb1}--\ref{cleb2}). Substituting the Clebsch-Gordon coefficients into (\ref{deqs1}--\ref{deqs2}) yields explicit expressions for the expansion coefficients $\{ d_{-2}^{(0)},  d_{-1}^{(0)}, d_{-1}^{(1)}, d_{0}^{(2)}, d_{+1}^{(0)}, d_{+1}^{(1)}, d_{+2}^{(0)}  \}$, which are listed in \ref{appendix-mst}, (\ref{d-explicit-1}--\ref{d-explicit-2}). 

To compute the spheroidal harmonics explicitly we require expressions for the spin-weighted spherical harmonics ${}_{-2}Y_l^2(x)$, where $x = \cos \theta$. These may be found by acting on spherical harmonics of spin-weight zero, ${}_{0}Y_l^0(x) \equiv \sqrt{\frac{2l+1}{4 \pi}} \, P_l(x)$, 
with ladder operators \cite{Goldberg-1967}. The spin-weight is lowered with the operator $\check{\delta}$, and the azimuthal number is raised with $L^+$. These operators are defined by
\begin{eqnarray}
\fl\quad \check{\delta} \, {}_sY_l^m(x) = \left( \sqrt{1-x^2} \, \partial_x - \frac{m + s x}{\sqrt{1-x^2}} \right) {}_sY_l^m(x) = -\sqrt{(l+s)(l-s+1)} \, {}_{s-1}Y_l^m(x), \\
\fl\quad  L^+ \, {}_sY_l^m(x) = -\left(\sqrt{1-x^2} \, \partial_x + \frac{s + mx}{\sqrt{1-x^2}} \right) {}_sY_l^m(x) = \sqrt{(l-m)(l+m+1)} \, {}_sY_l^{m+1}(x).
\end{eqnarray}
Here, $\partial_x$ is shorthand for the partial derivative with respect to $x = \cos\theta$. 
By acting with $\check{\delta} L^+ \check{\delta} L^+$ on ${}_{0}Y_l^0(x)$, it is straightforward to show that the spin-weighted harmonics in (\ref{fg1}) and (\ref{fg2}) can be written
\begin{eqnarray}
{}_{-2}Y_l^2 (x) &= \sqrt{\frac{2l+1}{4 \pi}} \frac{ \diffop \, P_l(x) }{ (l-1)l(l+1)(l+2) }  ,  \\ 
  \diffop P_l(x)  &= (1+x)^2 \, \partial_x (1-x) \partial_x \partial_x (1-x) \partial_x P_l(x)  \label{spin-2}.
\end{eqnarray}
Their values in the forward and backward directions are particularly simple,
\begin{equation}
{}_{-2}Y_l^2(x=1) = \sqrt{\frac{2l + 1}{4 \pi}} \, , \quad \quad {}_{-2}Y_l^2(x=-1) = 0 .
\end{equation}
This implies that the values of the spheroidal harmonics in the forward direction are
\begin{equation}
{}_{-2}S_l^m(x=1)  =  \sqrt{\frac{2 l + 1}{4 \pi}} \left( 1 + \mathcal{S}_1 z  + \mathcal{S}_2 z^2 + \mathcal{O}(z^3) \right)  \label{forward-direction}
\end{equation}
where
\begin{eqnarray}
\mathcal{S}_1 &= \frac{8}{(l+1)^2 l^2} , \\
\mathcal{S}_2 &= \frac{-3375}{(2l+3)^2(2l-1)^2} + \frac{16}{l^2(l+1)^2} + \frac{192 (l^2 + l + 1)^2}{(l+1)^4 l^4} - \frac{32}{(l+1)^4 l^4} .
\end{eqnarray}



\subsection{\label{subsec-amplitudes}Scattering Amplitudes}

\subsubsection{Helicity-conserving amplitude}
Let us first consider the helicity-conserving amplitude $f$ defined in (\ref{fg1}). To begin, we note that the sum of the positive and negative-parity phase terms can be written
\begin{equation}
\frac{1}{2}\left( e^{2i\delta^-_{lm\omega}} +  e^{2i\delta^+_{lm\omega}}  \right) = e^{i\chi} \, \frac{\Gamma(l - 1 - i\eps)}{\Gamma(l + 3 + i\eps)} \frac{\Gamma(l + 3)}{\Gamma(l - 1)} \left[ 1 +  \betco \eps^2  + \mathcal{O}(\eps^3) \right] , \label{plus-minus}
\end{equation}
where $
e^{i\chi} =  e^{2i\eps \ln |2\eps|} e^{-i \eps \kappa}
$
and the second-order coefficient is
\begin{equation}
\betco =  - i \pi \, \frac{\omega}{|\omega|} \left( \frac{\nu - l}{\epsilon^2} \right) - i \frac{m \astar}{l(l+1)} + \frac{2}{l(l+1)} - \frac{15}{(2l+3)(2l-1)} .   
\end{equation}
Inserting the expansions of the spheroidal harmonics (\ref{S-expansion}, \ref{forward-direction}), and using (\ref{plus-minus}), we may write the $f$ amplitude as
\begin{equation}
f(x) = \frac{e^{i\chi}}{2 i \omega} \diffop F(x) ,  
\end{equation}
where
\begin{equation}
\fl\quad\quad F(x) = \sum_{l=2}^{\infty} (2l+1) \frac{\Gamma(l - 1 - i\epsilon)}{\Gamma(l+3+i\epsilon)} \left( 1 + \betco \eps^2 + \mathcal{O}(\eps^3) \right) \left(1 + \mathcal{S}_1 \aw +  \mathcal{S}_1 \aw^2 + \mathcal{O}(\aw^3) \right)  V_l , \label{f-F}
\end{equation}
and
\begin{eqnarray}
\fl \quad\quad V_l =&  \sqrt{ \frac{2l-3}{2l+1} }  \frac{(l+1)(l+2)}{(l-3)(l-2)} b_{l-2}^{(l)} P_{l-2}(x) +  \sqrt{ \frac{2l-1}{2l+1} } \frac{(l+2)}{(l-2)} b_{l-1}^{(l)} P_{l-1}(x) + b_l^{(l)} P_l(x) \nn  \\
 &+  \sqrt{ \frac{2l+3}{2l+1} }  \frac{(l-1)}{(l+3)} b_{l+1}^{(l)} P_{l+1}(x) + \sqrt{ \frac{2l+5}{2l+1} } \frac{(l-1)l}{(l+3)(l+4)} b_{l+2}^{(l)} P_{l+2}(x)
\end{eqnarray}
and $b_k^{(l)}$ are the expansion coefficients defined in (\ref{S-expansion}).

To compute the higher-order terms in the sum $F(x)$, we substitute in the explicit forms for the coefficients $b_k^{(l)}$ calculated in Section \ref{sec-spheroidal}. Next, we rewrite the sum so all terms have a common factor of $P_l$. To demonstrate this process, let us begin by considering just the linear term in $\omega$, which we denote $F_{(\omega)}$. We find
\begin{eqnarray}
\fl F_{(\omega)}(x)  &\fl\quad\quad\quad=  a \omega \sum_{l = 2}^{\infty} (2l+1) \frac{\Gamma(l - 1 - i\epsilon)}{\Gamma(l+3+i\epsilon)} \left( \sqrt{\frac{2l-1}{2l+1}} \frac{(l+2)}{(l-2)} d_{-1}^{(0)}(l) P_{l-1}(x) +  \mathcal{S}_1 P_l(x) \right. \nn \\ & \quad\quad\quad\quad\quad\quad\quad\quad\quad\quad\quad\quad \left. + \sqrt{\frac{2l+3}{2l+1}} \frac{(l-1)}{(l+3)} d_{+1}^{(0)}(l) P_{l+1}(x) \right) \nn \\ 
\fl &\fl\quad\quad\quad= a \omega \sum_{l = 2}^{\infty} (2l+1) \frac{\Gamma(l - 1 - i\epsilon)}{\Gamma(l+3+i\epsilon)} \left( \sqrt{\frac{2l+3}{2l+1}} d_{-1}^{(0)}(l+1) + \mathcal{S}_1 + \sqrt{\frac{2l-1}{2l+1}} d_{+1}^{(0)}(l-1) \right) P_l(x)  + \mathcal{O}(\omega^2) \label{F-linear-resum} \nn \\ 
\fl &\fl\quad\quad\quad= 0 + \mathcal{O}(\omega^2) .
\end{eqnarray}
It is straightforward to verify that the terms in parantheses on the second line cancel exactly, and so a linear-in-$\omega$ term in $F(x)$ is not present.

Note that to obtain $P_l(x)$ as a common factor in (\ref{F-linear-resum}) we redefined the summation variable (i.e. $l \rightarrow l \pm 1$). Care must be taken in this process, since it changes the lower limit of the sum. In the above, we added an `extra' $P_2(x)$ term to the series without consequence, because the coefficient $d_{+1}^{(0)}(l-1)$ is zero for $l=2$. We also neglected a $P_1(x)$ term (present in the top line but not subsequently). This is justified for our purposes because, to obtain the final amplitude $f$, we act on $F$ with $\diffop$, and $\diffop P_1(x) = 0$. 

Let us now repeat this process and focus only on the part which is linear in $a$. This time, we will keep terms up to second order in $\omega$. Such terms arise from (i) the $\astar$-dependent part of the phase $\betco \eps^2$, and (ii) the linear term $\mathcal{S}_1$ coupled to an $\eps$ term arising from the effect of redefining the summation variable. We may split the result into two parts: a sum which turns out to be zero, and an $l=2$ term which arises from the change of summation variable. That is,
\begin{equation}
F(x) \approx F_{0}(x) + \aw \eps \left(  F_{1}^{(\Sigma)} +   F_{1}^{(l=2)}  \right) + \mathcal{O}(\aw^2, \eps^2) + \mathcal{O}(\omega^3) .
\end{equation}
The lowest-order term
\begin{equation} 
F_{0} = \sum_{l=2}^\infty (2l+1) \frac{\Gamma(l-1-i\eps)}{\Gamma(l+3+i\eps)} P_l(x) \label{F0} 
\end{equation}
was defined and summed in \cite{Dolan-2008}. The polarizing terms are 
\begin{eqnarray}
\fl F_{1}^{(\Sigma)} (x) = i \sum_{l=2}^\infty  (2l+1)  \frac{\Gamma(l - 1 - i\epsilon)}{\Gamma(l + 3 + i\epsilon)} \times \nn \\ 
\fl \quad  \quad   \quad   \quad     \left( \frac{-2(l+1)}{(l-1)(l+3)} \sqrt{\frac{(2l+3)}{(2l+1)}} d_{-1}^{(0)}(l+1) - \frac{4}{l(l+1)}  + \frac{2l}{(l-2)(l+2)} \sqrt{\frac{(2l-1)}{(2l+1)}} d_{+1}^{(0)}(l-1)  \right) P_l(x)  \nn \\
\fl \quad \quad \quad= 0 
\end{eqnarray}
and 
\begin{eqnarray}
F_{1}^{(l=2)} (x) &= -2i \frac{\Gamma(1-i\epsilon)}{\Gamma(5+i\epsilon)} P_2(x) .  \label{F1-l2}
\end{eqnarray}
The term $\aw \epsilon F_{1}^{(l=2)} (x)$  is responsible for the lowest-order polarizing effect. It leads to a term in the scattering amplitude which depends on the sign of $\omega$. That is, $f \approx f_{\text{Schw}} + f_{\text{pol}}$ where
\begin{eqnarray}
f_{\text{pol}} = -2M a\omega e^{i \chi} \left[ \frac{1}{2}(1+x) \right]^2 \frac{\Gamma(1-i\epsilon)}{\Gamma(1+i\epsilon)} \left( 1 + \mathcal{O}(\epsilon) \right) .
\end{eqnarray}
Since $f_{\text{pol}}$ is in phase with $f_{\text{Schw}}$, this gives a first-order contribution to the cross section,
\begin{eqnarray}
2 \left| f_{\text{Schw}}^\ast f_{\text{pol}}   \right|
= -4 a \omega M^2 \frac{ \cos^8(\theta/2) }{ \sin^2(\theta/2)} .
\end{eqnarray}

\subsubsection{Helicity-reversing amplitude}
We now repeat the analysis for the helicity-reversing amplitude, $g(\theta)$. To simplify matters, we only expand the relevant terms to first-order to recover the polarizing correction.
Let us begin by writing the amplitude as
\begin{equation}
g(\theta) = \frac{e^{i\chi}}{2 i \omega} \diffopm G(x),  \quad \text{where} \quad G(x) = \sum_{l=2}^\infty (-1)^l W_l V_l (1 + \mathcal{S}_1 \aw + \mathcal{O}(\aw^2) ) , 
\end{equation}
and
\begin{eqnarray}
\fl W_l &= 6i \eps  \frac{\Gamma(l - 1 - i \epsilon)}{\Gamma(l + 3 + i\epsilon)} \, \frac{\Gamma(l - 1)}{\Gamma(l + 3)}  \left( 1 + \frac{2 \astar m}{l(l+1)} \epsilon + \mathcal{O}(\epsilon^2) \right)  , \\
\fl V_l &= P_l(-x) + \aw \sqrt{\frac{2l-1}{2l+1}}\frac{l+2}{l-2} d_{-1}^{(0)}(l) P_{l-1}(-x) + \aw  \sqrt{\frac{2l+3}{2l+1}} \frac{l-1}{l+3}  d_{+1}^{(0)}(l) P_{l+1}(-x) + \mathcal{O}(z^2) .
\end{eqnarray}
As before, we may move terms of the series up or down ($l \rightarrow l \pm 1$) to obtain a common factor of $P_l(-x)$. As before, we find that, to lowest-order, the terms in the sum cancel out, leaving only an $l=2$ term which arises from redefining the summation variable. That is, 
\begin{equation}
G(x) \approx G_{0}(x) + \aw \eps \left( G_{1}^{(\Sigma)} +  G_{1}^{(l=2)} \right) + \mathcal{O}(\aw^2, \eps^2) + \mathcal{O}(\omega^3),
\end{equation}
and
\begin{eqnarray}
\fl G_{1}^{(\Sigma)} (x) &= 6 i \sum_{l=2}^\infty (-1)^l \frac{\Gamma(l - 1 - i\eps)}{\Gamma(l + 3 + i\eps)} \, \frac{\Gamma(l-1)}{\Gamma(l+3)} \left( (2l+1) \frac{16}{l(l+1)} - \frac{2(l+2)^2}{l^2} + \frac{2(l-1)^2}{(l+1)^2} \right) P_l(-x)  \nn \\
\fl &= 0, \\  
\fl G_{1}^{(l=2)} (x) &= 2 i \frac{\Gamma(1 - i\epsilon)}{\Gamma(5 + i\epsilon)} \, P_{2}(-x) .
\end{eqnarray}
The $z \epsilon G_{1}^{(l=2)} (x) $ term is responsible for the lowest-order polarizing effect. It leads to a term in the scattering amplitude which depends on the sign of $\omega$. That is, $g \approx g_{\text{Schw.}} + g_{\text{pol}},$ where
\begin{eqnarray}
g_{\text{pol}} = 2M a\omega e^{i \chi} \left[ \frac{1}{2}(1-x) \right]^2 \frac{\Gamma(1-i\epsilon)}{\Gamma(1+i\epsilon)} \left( 1 + \mathcal{O}(\epsilon) \right) .
\end{eqnarray}
The first-order contribution to the helicity-reversal cross section is
\begin{eqnarray}
2 \left| g_{\text{Schw}}^\ast g_{\text{pol}}   \right|
= 4 a \omega M^2  \sin^6(\theta/2).
\end{eqnarray}


\section{\label{sec:discussion}Discussion and Conclusion}
In the preceding sections we have shown that, in the low-frequency regime ($M|\omega| \ll 1$), the scattering cross section for a gravitational wave impinging along the axis of a rotating black hole is
\begin{equation}
M^{-2} \frac{d \sigma}{d \Omega} \approx \frac{\cos^8 (\theta / 2)}{\sin^4 (\theta/2)}  \left[ 1 - 4 a \omega \sin^2(\theta/2) \right]   +  \frac{\sin^8 (\theta / 2)}{\sin^4 (\theta / 2)}  \left[ 1 + 4 a \omega \sin^2(\theta/2) \right] . \label{csec-final}
\end{equation}
The first term in (\ref{csec-final}) arises from the helicity-preserving interaction; the second term arises from that part of the interaction that reverses the helicity of the incident wave.

The cross section (\ref{csec-final}) clearly depends on the sign of $\omega$, so scattering from a rotating black hole induces a partial polarization in an initially unpolarized beam. The total polarization is
\begin{equation}
\mathcal{P} \equiv \frac{ {\frac{ d\sig}{ d\Omega }}{(\omega>0)} - {\frac{d\sig}{d\Omega}}{(\omega<0)} }
{{\frac{ d\sig}{ d\Omega }}{(\omega>0)} + {\frac{d\sig}{d\Omega}}{(\omega<0)}  } = - 4 a |\omega| \sin^2(\theta/2) \left( \frac{\cos^8(\theta/2) - \sin^8(\theta/2)}{\cos^8(\theta/2) + \sin^8(\theta/2)} \right)  .
\label{pol-final}
\end{equation}
In the small-angle limit, the polarization is exactly half that predicted by Guadagnini and Barbieri \cite{Guadagnini-2002, Barbieri-2004, Barbieri-2005, Guadagnini-2008} for scattering from classical rotating matter (eq. \ref{pol-guadagnini}). 

There are a number of possible extensions of this work. Firstly, the analysis could be extended to encompass waves approaching at arbitrary angles of incidence with respect to the rotation axis. The relevant partial wave formulae, given in \cite{Futterman-1988}, involve an additional sum over the azimuthal numbers $m$. We would expect to find maximal polarization for waves impinging along the axis of rotation, and zero polarization for waves approaching in the equatorial plane. 
Secondly, the analysis could be repeated to compute the polarization of electromagnetic waves. Whilst we believe that the main features of this analysis follow through without difficulty, it remains to be checked. 
Finally, in a complementary work \cite{Dolan-2008b}, numerical methods have been employed to compute cross sections for arbitrary wavelengths. Excellent agreement between numerical and analytical results is found in the low-frequency regime. In addition, interesting higher-order effects are observed for general $M\omega$, such as interference fringes and glory halos.

\ack
Thanks to Bahram Mashhoon and Vitor Cardoso for thought-provoking correspondence, and to Marc Casals and Barry Wardell for proof-reading the manuscript and for helpful suggestions. Financial support from the Irish Research Council for Science, Engineering and Technology (IRCSET) is gratefully acknowledged.


\appendix

\section{Coefficients\label{appendix-mst}}
The coefficients $a_n^{\nu}$ appearing in the MST formalism (\ref{Am-eq}--\ref{Ap-eq}) are 
\begin{eqnarray}
\fl a_{-2}^\nu &= - \frac{ (l-1+s)^2 (l+s)^2 \left[ (l - 1)\kappa - i m \astar \right] \left[ l\kappa - i m \astar \right]}{4(l-1)l^2(2l-1)^2(2l+1)} \, \eps^2 + \mathcal{O}(\eps^3), \label{a-nu-m2} \\
\fl a_{-1}^\nu &= i \frac{(l+s)^2 \left[l\kappa - i m \astar\right]}{2l^2(2l+1)} \, \eps - \frac{(l+s)^2}{2l^2(2l+1)} \left[ 1 + i \frac{ l \kappa - i m \astar }{(l-1) l^2 (l+1)} m \astar s^2 \right] \eps^2 + \mathcal{O}(\eps^3) , \\
\fl a_0^\nu &= 1, \\
\fl a_1^\nu &= i \frac{(l+1-s)^2 \left[(l+1) \kappa + i m \astar\right]}{2(l+1)^2(2l+1)} \, \eps + \frac{(l+1-s)^2}{2(l+1)^2(2l+1)} \left[1 - i \frac{(l+1)\kappa + i m \astar}{l(l+1)^2(l+2)} m \astar s^2 \right] \eps^2 , \\
\fl a_2^\nu &= - \frac{(l+1-s)^2(l+2-s)^2 \left[(l+1)\kappa + i m \astar\right] \left[(l+2)\kappa + i m \astar \right]}{4 (l+1)^2 (l+2) (2l+1) (2l+3)^2} \, \eps^2 + \mathcal{O}(\eps^3)  \label{a-nu-2} .
\end{eqnarray}

With the Cordon-Shortley phase convention, the Clebsch-Gordan coefficients appearing in (\ref{deqs1}--\ref{deqs2}) are
\begin{eqnarray}
\left< l, 1, 2, 0 | l + 1, 2 \right> &= \left[ (l-1)(l+3)/(2l+1)(l+1) \right]^{1/2} , \label{cleb1} \\
\left< l, 1, 2, 0 | l , 2 \right> &=  \left[4 / l(l+1) \right]^{1/2},  \\
\left< l, 1, 2, 0 | l - 1, 2 \right> &=  -\left[(l-2)(l+2) / (2l+1)l \right]^{1/2}  ,
\end{eqnarray}
and
\begin{eqnarray}
\left< l, 2, 2, 0 | l+2, 2 \right> &= \left[ 6 \frac{(2l)!}{(2l+4)!} \frac{(l+4)!}{(l+2)!} \frac{(l!}{(l-2)!} \right]^{1/2} ,\\
\left< l, 2, 2, 0 | l+1, 2 \right> &= \left[ \frac{12}{(2l+1)} \frac{(l-1)(l+3)}{l (l+1)(l+2)}  \right]^{1/2} ,\\
\left< l, 2, 2, 0 | l, 2 \right> &=  \frac{-2(l-3)(l+4)}{\left[ (2l-1)(2l)(2l+2)(2l+3) \right]^{1/2} } ,  \\
\left< l, 2, 2, 0 | l-1, 2 \right> &= - \left[ \frac{12}{(2l+1)} \frac{(l-2)(l+2)}{(l-1)l(l+1)} \right]^{1/2}, \\ 
\left< l, 2, 2, 0 | l-2, 2 \right> &= \left[ 6 \frac{(2l-3)!}{(2l+1)!} \frac{(l+2)!}{l!} \frac{(l-2)!}{(l-4)!} \right]^{1/2}  .\label{cleb2}
\end{eqnarray}

The expansion coefficients $\{ d_{-2}^{(0)},  d_{-1}^{(0)}, d_{-1}^{(1)}, d_{0}^{(2)}, d_{+1}^{(0)}, d_{+1}^{(1)}, d_{+2}^{(0)}  \}$ appearing in (\ref{bdef}) are
\begin{eqnarray}
d_{+1}^{(0)}(l) &= 2 \left[ (2l+1) (2l+3) \right]^{-1/2}  \frac{(l-1)(l+3)}{(l+1)^2} , \label{d-explicit-1}  \\
d_{-1}^{(0)}(l)  &= -2 \left[ (2l+1) (2l-1) \right]^{-1/2} \frac{(l-2)(l+2)}{l^2} , \\
d_{+1}^{(1)}(l) &= 4 \left[ (2l+1) (2l+3) \right]^{-1/2} \frac{(l-1)(l+3)((l+1)^2 - 8)}{l(l+1)^4(l+2)} , \\
d_{-1}^{(1)}(l) &= -4 \left[ (2l+1) (2l-1) \right]^{-1/2} \frac{(l-2)(l+2)(l^2 - 8)}{(l-1)l^4(l+1)} , \\
d_{+2}^{(0)}(l) &= \frac{1}{2} \left[ (2l+1) (2l+5) \right]^{-1/2} \frac{ (l-1)l(l+3)(l+4)((l+9)}{(2l+3)^2 (l+1)^2 (l+2)} , \\
d_{-2}^{(0)}(l) &= -\frac{1}{2} \left[ (2l+1) (2l-3) \right]^{-1/2} \frac{(l-3)(l-2)(l+1)(l+2)(l-8)}{(2l-1)^2 (l-1) l^2} .  \label{d-explicit-2}
\end{eqnarray}

\section*{References}

\bibliographystyle{unsrt}


\end{document}